# Data acquisition and control system for neutron powder diffraction instrument

MW van der Watt*, A. Joel, A.P. van Dyk and AM Venter

NECSA Limited, PO Box 582, Pretoria, 0001, South Africa

* Corresponding author: e-mail: ndiff@aec.co.za

## Abstract

The development of control software for a neutron powder diffraction instrument at the Safari-1 research reactor of South Africa is reported. The software runs under Windows 2000 and incorporates PC-based cards for data acquisition and motion control. It has the capability to collect data from multiple detectors and controls three stepper motors. All stepper motors are equipped with encoders to ensure positional accuracy. The software drivers for detection and motion control are modular so as to enable their incorporation into other applications. Commands are performed in a batch format for the execution of command sequences without continuous input from the user. Certain limitations of the PC cards were overcome through the development of specialized drivers.

## 1. Introduction

New control software was developed for the upgraded powder diffraction instrument at the neutron diffraction facility of the Safari-1 research reactor. The instrument was upgraded from a single detector to a five-detector system. The software functionality was set to perform simple powder diffraction scans using the existing instrumentation for detection and motion control. A graphical user interface (GUI) has been created that enables the execution of multiple commands in a batch format as well as displaying the current status of the instrument. The data are written to file for off-line processing. The instrument description, user interface and software drivers are discussed.

## 2. Instrument description

The powder diffractometer has two rotating axes, the sample axis that manipulates the sample and the two-theta axis on which the detection system is moved around the sample. Stepper motors drive both axes in conjunction with encoders to ensure positional accuracy. The detection system consists of five detectors at a radius of 1m from the instrument centre. There is a 4° offset between detectors. Each detector is combined with a high transmission, $Gd_2O_3$ coated, Soller collimator that defines the diffracted beam divergence.

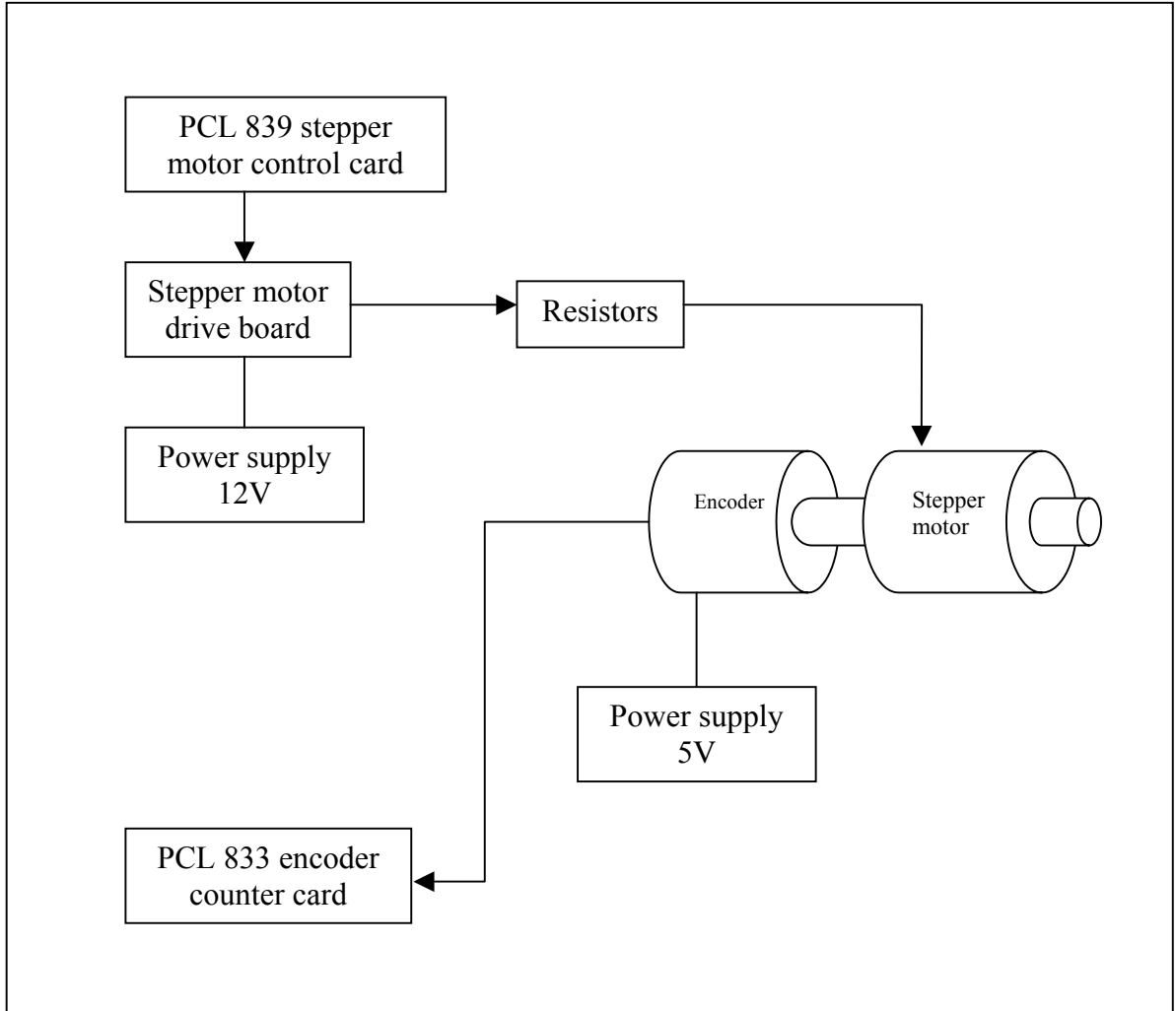

**Figure 1:** Schematic of motion control hardware (For simplicity only one stepper motor is shown).

Motion control consists of two parts, controlling the stepper motor movement and verifying its final position. Fig. 1 is a schematic of the motion control hardware. The stepper motor is controlled via the PCL-839 3-axis high-speed stepper motor control card from Advantech. The card communicates with a stepper motor drive board that supplies the power and switching of the stepper motor. The in-line resistors are used to adjust the voltage to suit the stepper motor specifications. The positional accuracy is verified with an incremental encoder mounted on the same axis as the stepper motor. For each revolution the encoder produces 1000 TTL-pulses that are counted by the PCL-833 3-axis encoder counter card. The positional accuracy is determined by comparing angular displacements of the stepper motor and the encoder.

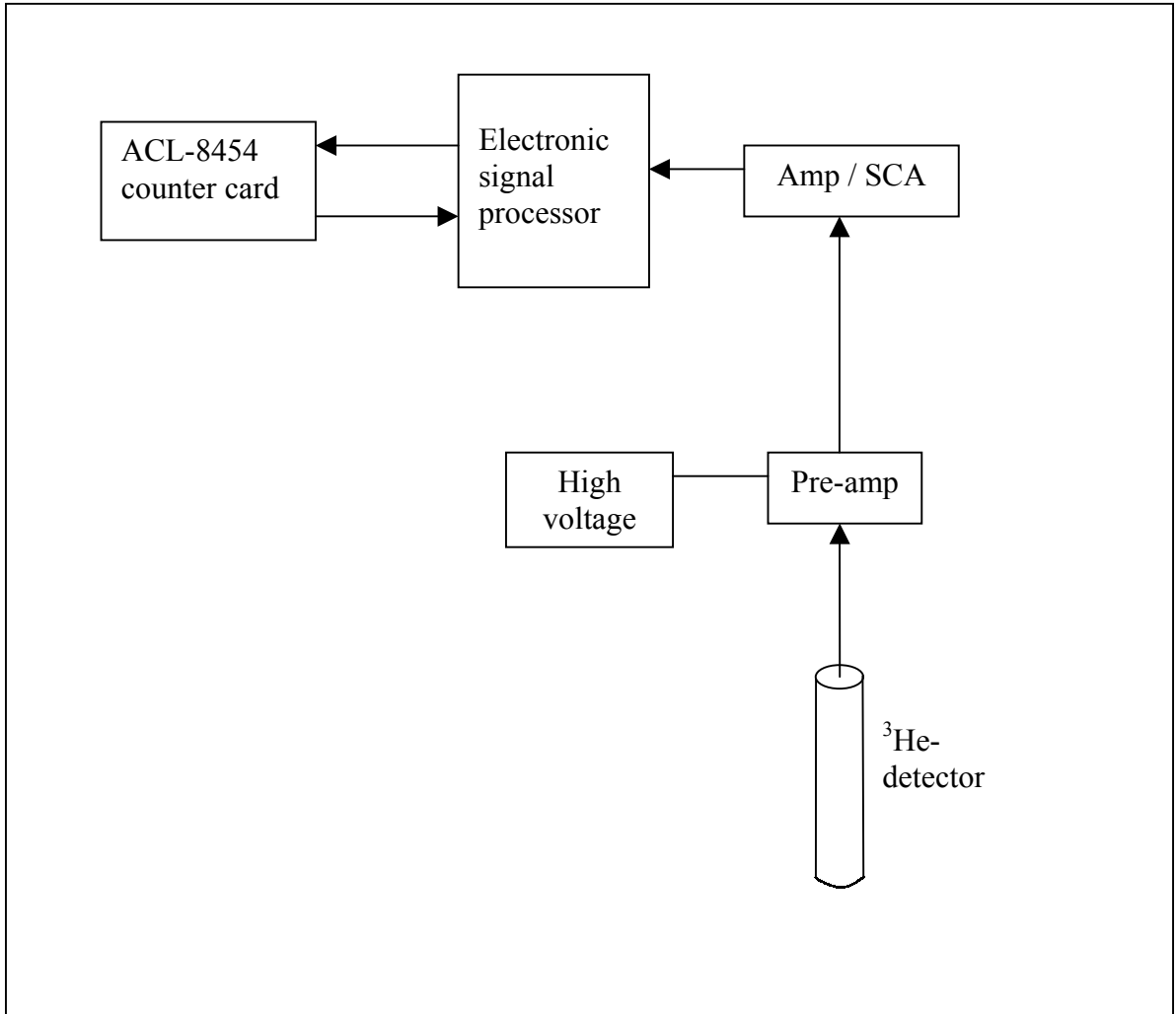

**Figure 2:** Schematic of detection hardware (For simplicity only one detector is shown).

An ACL-8454 counter/timer card from ADLink is used to count the events from the five neutron detectors and the in-line beam monitor. Fig. 2 shows the layout of the detection hardware. A pre-amplifier amplifies the signal from the $^3$He-detector, which is then processed by an Amp/SCA unit as TTL outputs. The signals are passed through an electronic signal processor that performs pulse shaping by 50$\Omega$ termination and a low-pass filter. The electronic signal processor is also used to switch between external neutron pulses and internal pulses generated on the counter card. These internal pulses are used to initialise the counter. The internal pulses proved invaluable in providing constant frequency pulses that were used for testing purposes during the development of the system. Further details are given in section 4.

**3. Graphical User Interface (GUI)**

The software was developed with C++ Builder and runs under the Microsoft® Windows 2000 operating system. In the following the layout of the GUI is discussed:

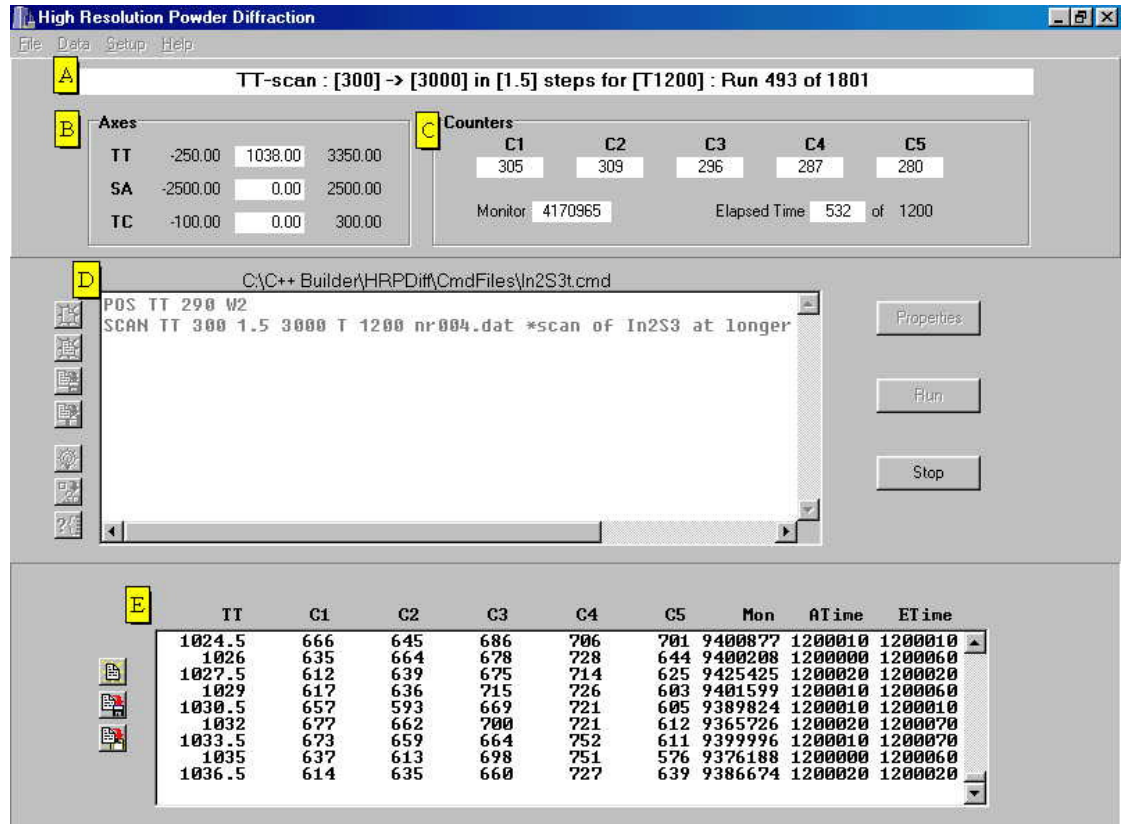

**Figure 3:** GUI of the control software.

The GUI is divided into five main areas as shown in Fig. 3:

"A": The current command being executed.

"B": The current status of the motion control with respect to lower limit, current position and upper limit.

"C": The counts registered by the five neutron detectors, the in-line beam monitor count and the elapsed time.

"D": The command editor where the user-entered commands are displayed and executed. To the left of the window are several speed buttons for various file functions. The three buttons to the right control most of the functionality of the program. The Properties button calls a pop-up dialog where general properties of the experimental environment are logged. The Run button sequentially executes each command entered in the command editor. The Stop-button acts as an emergency stop of the motion control and counting functions. Two software tests are run on every command before execution: Firstly, each command is verified for syntactical correctness. Secondly, the program steps through every command to verify that the axes positional limits are not exceeded. Error messages guide the user to correct faulty instructions.

"E": The data collected is displayed in tabular form.

## 4. Software drivers

A specialized driver was developed for the stepper motor control as the acceleration function of the PCL 839 card was too fast for the weight of the diffractometer. A smooth transition between the initial speed and the final speed is required as shown in Fig. 4.

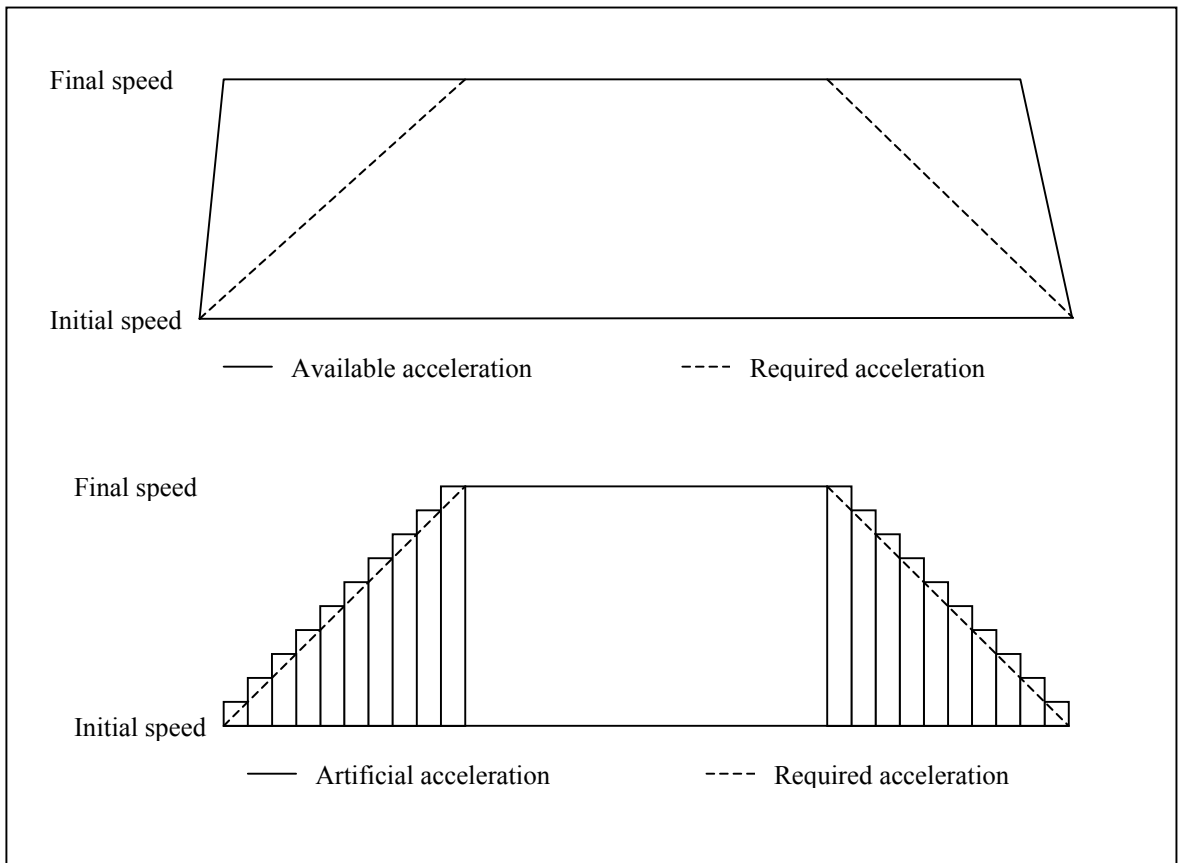

**Figure 4:** Artificial acceleration.

This has been accomplished artificially by advancing the stepper motor small distances and sequentially incrementing the final speed to the desired final speed. Only a few steps are shown for simplicity. The number and size of the steps may be changed to give any acceleration required. Deceleration is handled on the same principle.

A software driver was developed to address two limitations of the counter card functionality:

1. For timing purposes it is essential that all counters be initialised before a counting cycle begins. A counter is initialised when an input pulse is received. To guarantee that this condition is strenuously met, one pulse is supplied to each counter before a counting cycle begins via the electronic signal processor. This is done by momentarily switching between the external neutron pulses and internal pulses generated on the counter card. The switching is controlled by a switch signal sent by the software driver. Fig. 5 is a schematic of the process.

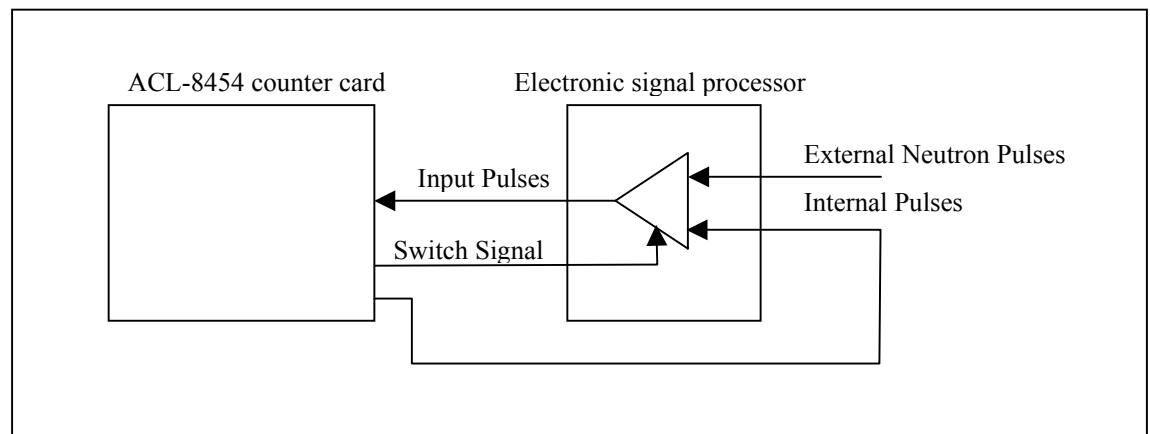

**Figure 5:** Schematic of switching between external and internal pulses.

2. The 16-bit counters proved inadequate as high count rates are experienced in the application. The functionality of the card is such that when a counter reaches its maximum value of 65536 it wraps around and starts counting from zero again. The software driver was modified to continuously test for this condition and keep track of how many times the counter has wrapped around. The true accumulated count can thus be determined. With these modifications all counting requirements were met.

**5. Future developments**

Extra functionality will be added to the software. These include automated calibration procedures, data processing and the introduction of a temperature controller to enable low temperature measurements in conjunction with a cryostat. The ISA-based PC-cards need to be updated to PCI-based formats. Due to the modular design of the software drivers, upgrades could be implemented without changes to the user program.

**6. Conclusion**

The program provides a sound platform to develop more sophisticated control software for the powder diffraction instrument. The modular design of the software drivers simplifies upgrades and can be used to create user group software for other applications.